\def\draft{1}
\def\llncs{0}

\def\anon{0}

\ifnum\draft=0

\fi

\ifnum\llncs=0
\documentclass[11pt]{article}
\usepackage{fullpage}
\else
\documentclass[orivec,runningheads,envcountsame,envcountsect]{llncs}
\usepackage{amssymb,amsmath,cite,url}
\fi

\usepackage{bm,bbm}

\newcommand{\remove}[1]{}
\newcommand{\ignore}[1]{}
\ifnum\llncs=0
\usepackage{amsthm}
\else
\fi

\ifnum\draft=1
\def\ShowAuthNotes{1}
\else
\def\ShowAuthNotes{0}
\fi

\usepackage{amsmath,amssymb}
\usepackage{url}
\usepackage{cite}
\usepackage{comment}
\usepackage[dvipsnames]{xcolor}
\usepackage{hyperref}
\hypersetup{colorlinks=true,
    allcolors=NavyBlue,
}

\definecolor{DarkBlue}{RGB}{0,0,150}
\definecolor{llg}{gray}{0.95}
\definecolor{lg}{gray}{0.85}

\ifnum\llncs=0

\newtheorem{assumption}{Assumption}

\newtheorem{property}{Property}[section]
\newtheorem{challenge}{Challenge}

\else
\spnewtheorem{prop}{Property}{\bfseries}{\itshape}
\spnewtheorem{fact}{Fact}{\bfseries}{\itshape}
\spnewtheorem{subclaim}{Claim}[theorem]{\bfseries}{\itshape}
\spnewtheorem{tlclaim}[theorem]{Claim}{\bfseries}{\itshape}
\spnewtheorem{assumption}{Assumption}{\bfseries}{\itshape}
\spnewtheorem{challenge}{Challenge}{\bfseries}{\itshape}
\fi

\ifnum\llncs=1

\else

\fi

\def\cA{{\cal A}}

\def\cD{{\cal D}}

\def\cS{{\cal S}}

\def\q2{\lfloor q/2 \rceil}

\newcommand{\mx}[1]{\mathbf{{#1}}}

\newcommand{\boldpar}[1]{\vspace{3pt}\par\noindent\textbf{#1}}

\ifnum\ShowAuthNotes=1
\newcommand{\authnote}[3]{\textcolor{#3}{[{\footnotesize {\bf #1:} { {#2}}}]}}
\else
\newcommand{\authnote}[3]{}
\fi

\newcommand{\absnewline}{\ifnum\llncs=1 \\ \fi}

\def\mgp[#1]{\mx{G}_{#1}}

\def\mgip[#1]{\mx{G}_{#1}^{-1}}

\def\mcit[#1]{\mci[#1]^T}
\def\mci[#1]{\mx{C}_{#1}}

\newcounter{hybridcount}
\newcounter{prevhybridcount}
\newcounter{nexthybridcount}

\def\beginM{\left[\begin{matrix}}
\def\endM{\end{matrix}\right]}

\def\m0{\mx{0}}

\usepackage{epigraph}
\usepackage[inline]{enumitem}
\title{Research Directions for Verifiable Crypto-Physically Secure TEEs\\
\vspace{5mm}
\texttt{draft}%
\ifnum\llncs=1
\ifnum\anon=0
\fi
\fi
}

\ifnum\llncs=1
\author{Sylvain Bellemare}
\institute{IC3, Cornell Tech, New York}
\date{}
\else
\ifnum\anon=1
\author{}
\else
\author{Sylvain Bellemare\thanks{\texttt{sbellemare@cornell.edu}} \\ \small{IC3, Cornell Tech}\thanks{Initiative for CryptoCurrencies and Contracts (\href{https://initc3.org}{initc3.org})}}
\fi

\ifnum\draft=1
\date{\today}
\else
\date{}
\fi

\fi

\ifnum\llncs=1
\renewcommand{\paragraph}{\boldpar}
\fi

\begin{document}

\maketitle

%\begin{quote}
%\emph{Information is not a disembodied abstract entity; it is always tied to a physical representation. It is represented by engraving on a stone tablet, a spin, a charge, a hole in a punched card, a mark on paper, or some other equivalent. This ties the handling of information to all the possibilities and restrictions of our real physical world, its laws of physics and its storehouse of available parts.}
%
%-- \textbf{Rolf Landauer}, in The physical nature of information
%\end{quote}

\epigraph{\textit{A castle is a very old example of a TEE.}}{C. Shepherd
    and K. Markantonakis \\ Trusted Execution Environments}

\epigraph{\textit{Information is not a disembodied abstract entity; it is always tied to a physical representation. %It is represented by engraving on a stone tablet, a spin, a charge, a hole in a punched card, a mark on paper, or some other equivalent. %This ties the handling of information to all the possibilities and restrictions of our real physical world, its laws of physics and its storehouse of available parts.
}}{Rolf Landauer \\ The physical nature of information}

\begin{abstract}
A niche corner of the Web3 world is increasingly making use of hardware-based Trusted
Execution Environments (TEEs) to build decentralized infrastructure. One of the
motivations to use TEEs is to go beyond the current performance limitations of
cryptography-based alternatives such as zero-knowledge proofs (ZKP), fully homomorphic
encryption (FHE), and multi-party computation (MPC).
%Thus, in the context of Web3, TEEs enable applications which may not be possible otherwise.
Despite their appealing advantages, current TEEs suffer from serious limitations as they
are not secure against physical attacks, and their attestation mechanism is rooted in the chip
manufacturer's trust. As a result, Web3 applications have to rely on cloud infrastruture
to act as trusted guardians of hardware-based TEEs and have to accept to trust
chip manufacturers. This work aims at exploring how we could potentially architect
and implement chips that would be secure against physical attacks and would not require
putting trust in chip manufacturers. One goal of this work is to motivate the Web3
movement to acknowledge and leverage the substantial amount of relevant hardware
research that already exists. In brief, a combination of:
\begin{enumerate*}[label = (\arabic*)]
    \item physical unclonable functions (PUFs) to secure the root-of-trust;
    \item masking and redundancy techniques to secure computations;
    \item open source hardware and imaging techniques to verify that a chip matches its expected design;
\end{enumerate*}
can help move towards attesting that a given TEE can be trusted without the need to
trust a cloud provider and a chip manufacturer.

\end{abstract}

\ifnum\llncs=0

\newpage
\tableofcontents

\newpage
\fi

\def\bhrd{\mathrm{BHRD}}
\def\efi{\mathrm{EFI}}
\def\wtW{\widetilde{W}}

\newcommand{\rdec}{\cA}
\newcommand{\sdec}{\cS}
\newcommand{\distn}{\cD}

\newcommand{\pieq}{\Pi_{\mathrm{EQ}}}

\section{Introduction}
This is an initiative to spark research to explore how we could develop a secure chip for TEEs (Trusted Execution Environments) that would ultimately be secure because of cryptophysics (i.e.\ physics, mathematics \& crytpography), rather than economics. This work is aimed at the Web3 (aka ``crypto'' \& blockchain) communities, who are increasingly using TEEs in various protocols to potentially secure substantial amounts of money, and/or to secure the privacy of users. We also hope to motivate more collaboration between the Web3 and the hardware communities, and thus, this work may also be interesting for some hardware communities, especially those interested in hardware-intrinsic security, and open source hardware.

% TODO
%A note about the thinking methodology \ldots on the role of scientific thought, separation of concerns
%\cite{dijkstra1982}.

%Side-channel and physical attacks cannot be prevented as of today. Making the cost of a chip attack expensive is the only current known defense mechanism \cite{cryptoeprint:2016/086}. Thus, TEEs are ultimately only secure through economics. The chip design should be open source, and its physical implementation should be verifiable, meaning that it should match the open source design. Moreover, the root of trust (embedded secret key) should be proven to have not leaked during generation or manufacturing. Thus, the hope and vision is to develop an open source and verifiable TEE chip that is secure through physics and mathematics. For an example of a cryptographic protocol implementation that is secure through physics see \href{https://www.nature.com/articles/s41586-021-03998-y}{Experimental relativistic zero-knowledge proofs} by \emph{Alikhani et al} \cite{alikhani2021experimental}.

\subsection{The Problem TEEs aim to solve}
TEEs are an attempt to solve the \emph{secure remote computation} problem. Quoting\cite{cryptoeprint:2016/086}: %\emph{Victor Costan and Srinivas Devadas}
\begin{quote}
\emph{``Secure remote computation is the problem of executing software on a remote computer \textbf{owned and maintained by an untrusted party}, with some integrity and confidentiality guarantees''.}
\end{quote}
The problem could potentially be solved by combining proving systems (e.g.\ zkVMs) with FHE schemes, but despite progress being made in this direction, the performance of these systems limits the applications that can be implemented, and as a result, some Web3 projects are turning to TEEs, as they offer better performance. These projects are however facing a serious problem: Current commercial TEEs exclude side-channel and physical protection\cite{cryptoeprint:2016/086,hasp:Rabimba_2021,arxiv:schneider2022sok:hwtee}, which implies that attackers with physical access to the TEE could compromise both the integrity and confidentiality guarantees. In order to mitigate this limitation, some projects propose to restrict the TEE to be run in a cloud provider, thus making the cloud provider a trusted party, responsible to protect the TEE from malicious physical attackers\cite{tee:pros:flashbots}. Although reasonable, this approach runs counter with the decentralization goals of Web3. Recently, different teams from the Web3 world, called for the need to seek alternative solutions\cite{tee:beyond:flashbots,tee:autonomous-manifesto:poetic,tee::poetic-intents}.

%Note that the remote computer is said to be owned and maintained by an \emph{untrusted} party. Yet,
%Current TEEs, cannot handle side-channel and physical attacks such as chip attacks (see \hyperref[Physical-Attacks-on-Chips]{Physical Attacks on Chips}), which would allow an attacker to retrieve the root of trust (secret keys encoded in the hardware). Once an attacker knows the secret keys, it can emulate a TEE, and go through the attestation process unnoticed (e.g.~see Appendix A. Emulated Guard eXtensions in https://sgx.fail/paper).

%Is it even possible to build a chip that can handle physical attacks, such as those making use of Focus Ion Beam microscopes as mentioned in \href{https://eprint.iacr.org/2016/086}{Intel SGX Explained} (section 3.4.3), and \href{https://dl.acm.org/doi/10.1145/2508859.2516717}{Breaking and entering through the silicon}? One could argue that it's not possible in the classical setting, but may be possible in the quantum setting. Some argue that PUFs (Physical Unclonable Functions) cannot be broken and would therefore be a solution. However, there's plenty of research that focuses of breaking PUFs, and there's also active research in developping more secure PUFs. Hence, it seems reasonable to assume that PUFs are not an ultimate solution to chip attacks, although they do seem to be a major improvement. (See \hyperref[Root-of-Trust-with-PUFs]{Root of Trust with PUFs}.)

\subsection{The Misalignment Between TEEs and Web3}
Web3 thrives for decentralization meanwhile current Trusted Execution Environments imply a trusted manufacturer, and thus are misaligned with the decentralization direction of Web3. It is arguably more complicated than \emph{just} trusting the manufacturer though, as the physical implementation of the TEE must also be trusted, independently of whether the manufacturer is honest or not. One could also argue that Web3 and current TEEs have radically different adversarial models. At a high-level, Web3 is about trusting nothing and verifying everything, meanwhile current TEEs, require trust in the manufacturer at the very least, but also trust that those who have physical access to the TEEs will not attack it.

%\paragraph{Adversarial Model of Current TEEs}
\subsubsection{Adversarial Model of Current TEEs}
In this work we wish to focus on the fact that current TEEs, used by Web3 applications, are not secure against physical attacks, and that they assume an honest manufacturer. For detailed presentations of the threat model of Intel SGX/TDX and the likes, see\cite{arxiv:schneider2022sok:hwtee,cryptoeprint:2016/086,aktas2023intel,10.1145/3133956.3134098,7846943}. For a treatment of TEEs in the context of blockchain applications see\cite{hasp:Rabimba_2021}.

We should point out that some TEEs, used in smartphones, do aim to provide protection
against side-channel and fault injection attacks, but
have nevertheless been attacked\cite{shepherd2024trusted,10.1016/j.cose.2021.102471}. Moreover, these TEEs (e.g. Apple Secure Enclave) are not used for blockchain applications as far as we know.

%\paragraph{Trust Assumptions of Current TEEs}
\subsubsection{Trust Assumptions of Current TEEs}
In general, trusting current TEEs, such as Intel SGX, rests on the following assumptions:

\begin{assumption}[Secure Design]\label{assump:design}
Trust that the chip is \textbf{designed} and secured as per the claims of the chip maker.\footnote{This trust is necessary because the designs of commercial TEEs are not open source, and despite academic efforts such as\cite{cryptoeprint:2016/086} to explain the design, we cannot know all the details of the design, and need to trust the chip company.}     
\end{assumption}

\begin{assumption}[Manufactured as per the Secure Design]\label{assump:fab}
Trust that the chip is \textbf{manufactured} as per the secure design, as claimed by the chip maker.\footnote{Note that design and implementation flaws can be fixed and can happen whether the design is open source or not, whether the supply chain is correct, etc. Hence, design and implementation bugs can be treated separately. It could be argued that an open source hardware design may benefit from a broader community and overtime will contain less bugs than a closed source design\cite{simplecrypto}.}
\end{assumption}

\begin{assumption}[Secure Root-of-Trust Generation]\label{assump:rot-gen}
Trust that the \textbf{root-of-trust} is not leaked during the manufacturing process. This means trusting that the manufacturer, or any other entity, have no knowledge of the root-of-trust.
\end{assumption}

\begin{assumption}[Secure Root-of-Trust Post-Fabrication]\label{assump:rot-sec}
Trust that the \textbf{root-of-trust} cannot be extracted out ``cheaply'' or ``easily'' by an attacker who has physical access to the chip.
\end{assumption}

\begin{assumption}[Secure Computations]\label{assump:sec-comp}
Trust that computations, executed in enclaves, do not leak confidential data, under side-channel attacks such as power analysis.
\end{assumption}

\begin{assumption}[Unforgeable Attestations]\label{assump:attest}
    Trust that \textbf{attestations} cannot be forged, in one way or another. This implies multiple assumptions:
    \begin{enumerate*}[label = (\arabic*)]
        \item trusting that attestation keys are protected against side-channel (e.g.\ power analysis) or physical attacks
        \item trusting that attestation keys are not known to the chip manufacturer.\footnote{See \href{https://www.rfc-editor.org/rfc/rfc9334.html\#name-security-considerations}{RFC 9334 (section 12)}\cite{rfc9334} for security considerations when treating the topic of remote attestation.}, and
        \item trusting that the root-of-trust is secure and that deriving the attestation from the root-of-trust would not be feasible,
    \end{enumerate*}
%This may also involve having to trust the role of the manufacturer (e.g.~Intel SGX with EPID or DCAP).
\end{assumption}
\noindent
These assumptions are a source of troubles for the Web3 communities since they represent a misalignment with the goals of decentralization.

\paragraph{The Malicious Manufacturer Problem} The problem of having to trust a manufacturer, is not new, nor unique to Web3 concerns.
For instance, quoting\cite{tehranipoor2010hardware}:
\begin{quote}
\textit{While the economic benefits are clear and many of the manufacturers are honest, outsourcing
gives rise to a significant security threat. It takes only one malicious employee to
compromise the security of a component that may end up in numerous products.}

\textit{Simply stated the untrusted manufacturer problem occurs in two situations:}
\begin{itemize}
	\item \textit{an intellectual property (IP)-core designer inserts a secret functionality that deviates from the
	  	  declared specification of the module}
	\item \textit{a manufacturer modifies the functionality of the design during fabrication.}
\end{itemize}
\textit{Clearly, besides tampering the goal of the malicious manufacturer is to operate covertly. 
The ever increasing complexity of hardware and software designs makes it much easier
for untrusted manufacturers to achieve this goal.}
\end{quote}

\subsection{Restoring the Alignment Between TEEs and Web3}
Rather than working around the above assumptions, we propose to research how we could design and implement novel TEEs that would be aligned with the decentralization goals and adversarial model of Web3.

%\paragraph{Adversarial Model of Crypto-Physically Secure TEEs}
\subsubsection{Adversarial Model of Crypto-Physically Secure TEEs}
TEEs should be secure against side-channel and physical attacks, and consequently attackers with physical access are part of the threat model. Moreover, the attacker is considered to be unbounded with respect to resources and skills to perform side-channel and physical attacks. We assume the most sophisticated attacker, with the highest attack potential, as per the industry framework presented in\cite{adv-model:jil}. The manufacturer and its suppliers are considered to be adversaries, that can inject hardware trojans and deviate from the chip design that must be manufactured\cite{tehranipoor2010hardware}. For a classificaiton of the types of attacks we are concerned with see\cite{9889394,acosta2017embedded,chowdhury2021physical,mangard2008power}.

\paragraph{A Note About Economic Security} Although the focus of this work is to research directions to develop crypto-physically secure TEEs, if possible it is likely to take considerable time, and in the meantime, it would be important to acknowledge the economical dimension of attacks and to be able to identify the precise costs to perform the attacks, in order to inform protocol designers, who can then introduce game theoretic aspects in their designs\cite{hasp:Rabimba_2021}.

%\paragraph{Challenges for Crypto-Physically Secure TEEs}
\subsubsection{Challenges for Crypto-Physically Secure TEEs}\label{sec:challenges}
To help orient the research work, we present the following challenges to help us think beyond the current TEEs, and to move towards novel TEEs that do not sacrifice Web3's decentralization efforts. These challenges reflect our view that we need to move towards TEEs which are secure through physics and cryptography.
\begin{challenge}[Open Source Hardware Design]\label{challenge:oshw}
As per Kerckhoffs' principle: ``It must not require secrecy, and it must be capable without inconvenience to fall into the enemy's hand''.\cite{kerckhoffs}
\end{challenge}

\begin{challenge}[Proof of Fabrication]\label{challenge:proof-of-fab}
The physical implementation must match its open source design, and it must be verifiable.
\end{challenge}

\begin{challenge}[Oblivious Root-of-Trust Generation]\label{challenge:ORoTG}
The root-of-trust must not be known to any party before, during or after generation time. The process used to generate the root-of-trust must not leak any information that could be used to reconstruct the root-of-trust.
\end{challenge}

\begin{challenge}[Oblivious Root-of-Trust Usage]\label{challenge:ORoTU}
Using the root-of-trust to support computations, such as signatures and key derivations, must not reveal any information that could be used to reconstruct the root-of-trust, or any other confidential data for which the security is rooted in the root-of-trust.
\end{challenge}

\begin{challenge}[Crypto-Physically Secure Computations]\label{challenge:cryptophysically-secure-tee}
The trusted execution environments must not leak confidential data, and must be protected against side-channel attacks\cite{9889394,acosta2017embedded,chowdhury2021physical,mangard2008power}.    
\end{challenge}

\begin{challenge}[Unforgeable Attestations]\label{challenge:unforgeable-attestations}
Attestations must be authentic and verifiable such that a verifier can link the report to a secure root-of-trust that is tied to the verified implementation, which corresponds to the expected open source hardware design. 
\end{challenge}

\subsection{Related Work}\label{sec:related-work}
The Secure Cryptographic Implementation Association\cite{simplecrypto} is actively working on developing open hardware that can withstand physical attacks such as side-channel and fault attacks. Their \href{https://www.simple-crypto.org/about/vision/}{vision} is that an \emph{``open approach to security can lead to a better evaluation of the worst-case security level that is targeted by cryptographic designs''}. OpenTitan\cite{opentitan:10106105,opentitan:10.1145/3649153.3649213,opentitan:10.1145/3690823}
is an open source hardware root-of-trust. The importance of verifiable hardware and a
novel verification technique, based on infra-red light, are discussed
in\cite{huang2023infraredinsituirisinspection} by Andrew `bunnie' Huang. The importance
of open source hardware and verification techniques is discussed in\cite{10224911}. The
open hardware landscape is surveyed in\cite{9548130}. For a very detailed and broad
survey of TEEs, see\cite{shepherd2024trusted}, which motivates the need for more secure
TEEs, including protection against physical attacks.

Sanctum\cite{sanctum:hw-ext:197162,sanctum:secure-boot:8429295} addresses software-based
side-channel attacks, such as cache timing attacks via minimal hardware extensions in a
first work. They also point out the difficulty of securing remote computations even in
the presence of strong economic incentives. In a second work, Sanctum leverages PUFs to provide a secure boot and remote attestation. The threat model assumes an honest-but-curious manufacturer, meaning that the manufacturer may attempt to read the root-of-trust, but will implement the chip according to its expected design. In our case the manufacturer is assumed to be malicious, which means that we do not rely on the manufacturer to implement the chip as per the expected design.
Keystone enclave\cite{keystone:10.1145/3342195.3387532} is an open source software TEE
framework. It needs a hardware layer with a secure root-of-trust.
SpaceTEE\cite{DBLP:journals/corr/abs-1710-01430} uses nano-satellites (CubeSats) to
protect a hardware security module (SpaceHSM) against physical attacks.
%From our early review of SpaceTEE, it's not clear how the remote attestation mechanism works.

% Gassend, mentions certified execution and manufacturer resistant.

Motivation to put more research efforts into implementing roots of trust with emerging
hardware technologies, such as resistive memories and flexible electronics is discussed
in\cite{hw-sec:10533875}. A brief survey of challenges in hardware security is presented
in\cite{hw:9453101}. In the Web3 context, motivaton for research to address the
current limitations of TEEs and their lack of decentralization has been mentioned
in\cite{os:enclave:keystone,tee:beyond:flashbots,tee:autonomous-manifesto:poetic,tee::poetic-intents}.
In addition to adressing the lack of decentralization with current
TEEs,\cite{hasp:Rabimba_2021} also points out at the lack of an economic security
model to work with when integrating TEEs in a blockchain context.
In zach's tech blog\cite{zachbe:build-chips}, the author motivates how the current
semiconductor startup ecosystem could benefit from a Silicon Venture Studio.

There's a lot of work that covers the topic of using physics to secure a system, such
as\cite{pappu:doi:10.1126/science.1074376,puf:10.1145/586110.586132,alikhani2021experimental,brassard2021relativity}.
\section{Ethics, Morality, and the Cypherpunks}
Is it a good idea to attempt to build a secure-through-physics TEE?\ Are there dangers?
Could it be misused? This section is meant to invite the community to reflect on the
ethical and moral implications of an eventual secure-through-physics TEE.\ In order to
guide this reflection, we propose a few works to reflect on, as starting points. We
don't claim that they're the only ones to reflect on, nor that they're the best ones.
We simply wish to encourage discussions on the ethical and moral implications of our
work. 

\begin{itemize}
    \item The Moral Character of Cryptographic Work by Phillip Rogaway\cite{208512},
    \item The battle for Ring Zero by Cory Doctorow\cite{RingZero}
    \item The Crypto Anarchist Manifesto by Timothy C. May\cite{cam}
    \item A Cypherpunk's Manifesto by Eric Hughes\cite{cpm}
    \item Trusted Execution Environments (Section 8.5) by Carlton Shepherd and Konstantinos Markantonakis\cite{shepherd2024trusted}
    %\item The Oracle, by Ari Juels\cite{}
\end{itemize}

\paragraph{The battle for Ring Zero} Cory Doctorow, in\cite{RingZero} points out:
\begin{quote}
\emph{But how can we trust those sealed, low-level controllers? What if manufacturers
--- like, say, Microsoft, a convicted criminal monopolist --- decides to use its
low-level controllers to block free and open OSes that compete with it? What if a
government secretly (or openly) orders a company to block privacy tools so that it can
spy on its population? What if the designers of the secure co-processor make a mistake
that allows criminals to hijack our devices and run code on them that, by design, we
cannot detect, inspect, or terminate? That is: to make our computers secure, we install
a cop-chip that determines what programs we can run and stop. To keep bad guys from
bypassing the cop-chip, we design our computer so it can't see what the cop-chip is
doing.\textbf{So what happens if the cop-chip is turned on us?}}
\end{quote}

%\input{background}
%\input{adverserial-model}
%\input{cypherpunk-chip}

% the challenges/pillars
\section{Oblivious Root-of-Trust Generation}\label{sec:rot-gen}
%Recalling the trust assumption \ref{assump:rot-gen}, we want to avoid the situation where a manufacturer, or any other entity would have knowledge of the root-of-trust.
How can we create a signing key (root-of-trust) in an oblivious way, meaning that the
key must not be known by any party when it is created? Physical unclonable functions
(PUFs) and True Random Number Generators are the most secure known methods to
intrinsically generate keys. TRNGs generate different keys each time they are invoked,
and consequently if used for a root-of-trust, the key must be stored in non-volatile
memory (e.g., fuses). As an aside, Apple's Secure Enclave uses a TRNG to establish a
root-of-trust\cite{apple-secure-enclave}.
Non-volatile memory elements, such as fuses, are prone to imaging
attacks\cite{10.1145/2508859.2516717}, and consequently we do not wish to use a TRNG
to generate the signing key (root-of-trust).
PUFs can be used to re-generate the same key each time the chip is powered up and thus
do not require non-volatile memory to store the key.

%\paragraph{Physically Unclonable Functions (PUFs)}
\subsection{Physically Unclonable Functions (PUFs)}
Current state-of-the-art chip fabrication, although extremely precise, is not precise enough to prevent unpredictable and uncontrollable physical variations between identical logical components. For instance, two identical transistors will react differently under the same voltage. Hence, millions of chips with the same identical design, will have different physical properties, which can be leveraged as sources of entropy. A physical unclonable function (PUF) is a circuit block in a chip that leverages these physical variations to derive the unique fingerprint of its chip. This fingerprint can be used to uniquely identify and authenticate a chip and also to generate a signing key (root-of-trust)\cite{puf:maes2013}. Quoting\cite{puf:10.1145/586110.586132}:
\begin{quote}
``We wish to implement a PUF in silicon so we can \textbf{identify and authenticate} a given \textbf{integrated circuit} (IC). By exploiting \textbf{statistical variations} in the delays of devices and wires within the IC, we create a \textbf{manufacturer resistant} PUF''.
\end{quote}
Physically unclonable functions (PUFs) are said to be manufacturer resistant because the manufacturer cannot predict nor control the physical variations from which entropy is derived. Furthermore, the manufacturer, or anyone, cannot ``clone'' a PUF such that it will behave the same way as the original one. Also, for most PUFs, probing the PUF is expected to destroy its entropy, and therefore its associated fingerprint or key. It's important to note that we wish to point to PUFs as a very good potential candidate solution for Challenge~\ref{challenge:ORoTG}, (that of intrinsically creating a signing key, such that the key cannot be observed by any entity), but more study and research is required to properly understand their security limits and vulnerabilities. Exploring the security limits of PUFs is outside the scope of this work, but we nevertheless present a brief and partial survey of attacks on PUFs in Section~\ref{sec:puf-attacks}. As it will be seen in Section~\ref{sec:puf-properties} on the core properties of PUFs, the PUF is not really a function since its output is not deterministic as it contains noise from the physical environment of the chip, such as heat, and voltage. For cases like generating a signing key, a deterministic response is required and PUF designs must include some post-processing on the response to clean out the noise. Section~\ref{sec:puf-noise-removal} provides a small introduction to the topic of stabilizing a PUF response. We now briefly discuss the two main types of PUFs: strong and weak PUFs, also named authentication and key generation PUFs, respectively. 

\paragraph{A Note about Strong PUFs versus Weak PUFs} The lierature distinguishes between two main types of PUFs, based on the size of their challenge-response pairs (CRPs) set. PUFs that have a very large set of CRPs are referred to as ``strong'' or authentication PUFs, whereas PUFs that have very few CRPs, are referred to as ``weak'' or key (generation) PUFs. Strong PUFs although theoretically promising\cite{puf:rom:ra:van2023theoretical} have been subject to multiple attacks, most notably machine learning attacks, and have yet to be shown to be secure\cite{ieee:puf:strong:9971721}. On the other hand, weak PUFs, used for key generation, are more secure as their small set of CRPs cannot be exploited by machine learning attacks, and are already in use commercially extensively, especially in IoT devices.
%TODO add citation about the security of weak PUFs

\subsubsection{PUF Properties}\label{sec:puf-properties}
Our goal here in reviewing the properties of PUFs is to show how they are a good fit to address the challenge of obliviously generating the root-of-trust (Challenge~\ref{challenge:ORoTG}). We'll use the definitions presented in\cite{puf:maes2010} almost verbatim as a basis. The authors first define a PUF as a physical challenge–response procedure, which implies that it is embedded in a physical device, has an input/output mechanism and is more general than a function, since a PUF can have multiple outputs for the same input\cite{puf:maes2013}. A PUF is denoted as
\begin{equation}
    \Pi: \mathcal{X} \rightarrow \mathcal{Y} : \Pi(x) = y.
\end{equation}
Using the above defintion, the authors in\cite{puf:maes2010} outline the following semi-formal properties (see\cite{puf:maes2010} for more details).

\begin{property}[Evaluatable]
    Given a {\normalfont{\textsf{PUF}}} $\Pi$ and a challenge $x$, it is \texttt{easy} to evaluate the response $y = \Pi(x)$.
    %(Hence, the root-of-trust can be efficiently generated.)
\end{property}

\begin{property}[Unique]
    {\normalfont{\textsf{PUF}}} $\Pi(x)$ contains \texttt{some} information about the identity of the physical entity embedding the {\normalfont{\textsf{PUF}}} $\Pi$.
    %(We can use a {\normalfont{\textsf{PUF}}} to uniquely identify a chip (TEE).)
\end{property}

\begin{property}[Reproducible]
    The response $y = \Pi(x)$ is reproducible up to a \texttt{small} error.
    %(We can re-generate the same signing key, repeatedly, when powering up a chip.)
\end{property}

\begin{property}[Unclonable]
    Given a {\normalfont{\textsf{PUF}}} $\Pi$, it is \texttt{hard} to construct a procedure $\Gamma \neq \Pi$ such that $\forall x \in \mathcal{X} : \Gamma(x) \approx \Pi(x)$ up to a \texttt{small error}.
    %(It is practically infeasible to build chips, for which the {\normalfont{\textsf{PUF}}} will yield identical signing keys. Moreover a mathematical clone should also be hard (infeasible) to build, thus preventing simulating the {\normalfont{\textsf{PUF}}}, and its associated signing key.)
\end{property}

%\begin{property}[Unpredictable]
%    Given only a set $\mathcal{Q} = \{(x_i, y_i = \Pi(x_i))\}$, it is \texttt{hard} to predict $y_c \approx \Pi(x_c)$ up to a \texttt{small error}, for $x_c$ a random challenge such that $(x_c, \cdot) \not\in \mathcal{Q}$.
%\end{property}
%
%\begin{property}[One-way]
%    Given only $y$ and $\Pi$, it is \texttt{hard} to find $x$ such that $\Pi(x) = y$.
%\end{property}

\begin{property}[Tamper evident]
    Altering the physical entity embedding $\Pi$ transforms $\Pi \rightarrow \Pi^\prime$ such that with high probability $\exists x \in \mathcal{X} : \Pi(x) \neq \Pi^\prime(x)$, not even up to a \texttt{small error}.
    %(Probing the PUF in a chip should change its entropy, effectibely destroying the signing key.)
\end{property}

\noindent
\paragraph{Evaluatable} The root-of-trust can be efficiently generated.

\noindent
\paragraph{Uniqueness} The root-of-trust will be unique to each chip, which will allow identification of the chip, via the public key associated with the signing key (root-of-trust).

\noindent
\paragraph{Reproducibility (Stability)} The root-of-trust can be re-generated repeatedly, across the lifetime of the chip. Recall that the root-of-trust is only present when the chip is powered up. Hence, each time the chip is powered up, the exact same signing key must be re-generated. This is not so simple since the response of a PUF is subject to environmental noise, and aging, and thus will vary. Post-processing steps are necessary to clean out the noise, via error correcting codes or other means. The literature also refers to this property as stability.

\noindent
\paragraph{Unclonability} It is practically infeasible to build chips, for which the PUF will yield identical signing keys. Moreover a mathematical clone should also be hard (infeasible) to build, thus preventing simulating the PUF, and its associated signing key. Quoting\cite{puf:maes2010}:
\begin{quote}
[\ldots] the hardness of cloning can be considered from a theoretical and a practical point of view. Practically, cloning can be very hard or infeasible. Demonstrating theoretical unclonability on the other hand is very difficult. The only known systems which can be proven to be theoretically unclonable are based on quantum physics.
\end{quote}

%It's important to note that the hardness of cloning a PUF is practical one, whereas theoretical hardness can be achived in with quantum systems. 

\noindent
\paragraph{Tamper evidence} Physically probing the PUF in a chip should change its entropy, effectively destroying the root-of-trust (signing key). Note that this is the ideal goal, but that there has been attacks in which the PUF is probed and yet not altered such that extracting its response (root-of-trust in our case) was possible\cite{delvaux2017security}. As mentioned earlier, PUFs are an active area of research, and more secure PUFs are constructed, but also attacks, thus moving the technology forward. See Section~\ref{sec:puf-attacks} for a partial survey of physical attacks on PUFs.

\subsubsection{Good and Bad Chaos for PUFs}
As mentioned, a PUF leverages the physical variations introduced during manufacturing.
When a chip is powered up it is also subject to variations like heat, which
will affect the behavior of the PUF, such that its response will not be deterministic.
A chip is also subject to aging, and overtime the response of a PUF will also change
because of degradation of its physical components. Hence, there's good and bad chaos for
reliable PUFs. In order to help us better understand PUFs, we'll review the three types
of physical variations in (CMOS) chips, based on the work
in\cite{puf:kim2010statistics}.

\paragraph{Process Manufacturing Variations} Two main types of variations are introduced
during the making of chips:
\begin{enumerate*}[label = (\arabic*)]
	\item \emph{variations in process parameters} for the deposition and/or diffusion of dopants
	\item \emph{variations in the dimensions of the transistors}, due to the limited resolution
 		of the photo-lithographic process.\cite{puf:kim2010statistics}.
\end{enumerate*}

\paragraph{Environmental Variations} A PUF is embedded in a chip, and its surrounding
environment can have an impact on its response. Factors like heat, voltage and noise
coupling can cause the response of a PUF to be different within relatively short amounts
of time. For example, the vibration of electrons will cause electronic noise, known as
Johnson–Nyquist noise, which will impact a PUF's response. Hence, even in perfectly
stable laboratory conditions, a PUF's response is not perfectly
reproducible.\cite{delvaux2017security}

\paragraph{Aging} Depending on how frequently a chip is used, its components will slowly
degrade and these physical changes will impact the PUF response.

\subsubsection{PUF Response Stabilization}\label{sec:puf-noise-removal}
As mentioned previously, chips are subject to unwanted variations which will destabilize
the PUF response, and thus require post-processing steps to re-generate the same
signing key (root-of-trust). Error correction
schemes\cite{DBLP:journals/tches/BatinaCHSS23,ieee:ecc:10027089}
are commonly used to recover the original response. Although reliable, error correction
schemes require helper data to be stored in non-volatile memory. To circumvent this
drawback, other techniques have been developed, such as temporal majority voting,
dark-bit masking, and burn-in enhancement. Many PUF designs combine temporal majority
voting with dark-bit masking to achieve better stability of the PUF
response\cite{chuang2020highly}. Active PUFs were also introduced
in\cite{chuang2020highly}, by Chuang, as a new method to stabilize the response of a
PUF.\ Such PUFs require an activation step after the fabrication of the circuit.
In\cite{chuang2020highly}, a metal oxide breakdown PUF is proposed, in which two
logically identical transistors are stressed with a high voltage to provoke the
formation of oxide breakdown. Depending on which transistor breaks down first, a bit of
$0$ or $1$ is set. Since thee oxide breakdown is permanent the PUF response becomes
stable. By building an array of transistor pairs, a stable bit string can be obtained,
and used to to generate a key. However, it was shown in\cite{puf:attack:10445170}, to be
vulnerable to imaging attacks, by using Voltage Contrast Scanning Electron Microscopy
(VC-SEM), to identify which transistor has oxide breakdown, and then reconstruct the
key, with very high accuracy.

%As introduced earlier, a PUF leverages the manufacturing variations of a chip as a
% source of entropy, which can be used to generate a signing key (root-of-trust). There
%are other sources of physical variations which are problematic to a PUF's design and
%implementation. These variations are grouped into two categories:
%\begin{enumerate*}[label = (\arabic*)]
%	\item environmental variations
%	\item aging
%\end{enumerate*}

%Error correction codes \cite{}

%\paragraph{Recent PUF Constructions}
\subsection{Recent PUF Constructions}\label{sec:recent-pufs}
To give some pointers to recent research work in developing PUFs, readers may
consult\cite{puf:recent:8751997,puf:recent:8244298,puf:recent:8310218,neopuf:9440114}
but also note that the security of these PUFs has been attacked
in\cite{puf:attack:10445170}. Future version of this work intend to add state-of-the-art
PUF constructions to this section.

%Generally speaking PUFs, take a challenge as input and output a response. A PUF can take many challenges or a single one. PUFs that have a very large set of challenge-response pairs (CRPs) are often referred to as strong PUFs or authentication PUFs, whereas PUFs that have a very small amount of CRPs are called weak PUFs, or key-generation PUFs. Most strong PUFs if not all, have been shown to be broken, and consequently it is not clear whether strong PUFs can be reliable. In the meantime, weak PUFs are very well studied and used commercially in millions of devices. Weak PUFs are nevertheless not perfect and just like strong PUFs, are an active area of research and development.

\subsection{Physical Attacks on PUFs}\label{sec:puf-attacks}
Future work should aim at documenting all known physical attacks, but for now we'll
summarize the work of Delvaux in\cite{delvaux2017security}, and also point to recent
work shown in\cite{puf:attack:10445170} and\cite{ieee:puf:strong:9971721}. One important
topic to cover is that of attacks to read volatile memory, and protection against such
attacks\cite{10323716}. Another important category of attacks targets the
post-processing step (fuzzy extraction), via side-channel and/or fault injection
attacks\cite{pehl2024design,9344071,tebelmann2017side,karakoyunlu2010differential}.

\subsubsection{A Note on Characterization Techniques for Semiconductors}
This is for attacks that aim at characterizing the critical parts of a chip (e.g. PUFs), when it is powered down, in order to derive a mathematical model to perform simulations to predict the behaviour of the chip when it is powered up. This is super crucial to the current security of PUFs, and more precisely to their mathematical unclonability. It’s currently infeasible to characterize a PUF for instance, because the characterization techniques are not advanced enough. That being said, it does not mean that characterization techniques will not improve, and consequently it may be wise to encourage the development of these techniques as it well help us better understand the physical limits of the security of PUFs. Related research can be consulted in\cite{img:kim2022expected,img:nakano2023molecular,img:stransky2023water,sim:samuel2023cell}.

%According to François-Xavier Standaert, the tamper resistance mentioned in the early papers on PUFs is not enough. Tamper resistance and passive/active attacks must be considered jointly and there aren’t convincing papers on the topic so far. For instance, many works on fuzzy extraction ignore the risk of leakage when performing the extraction process, and also don’t consider fault attacks. (We should consult with François-Xavier again on this topic, and also with other experts.)

%\subsection{Future Research Directions}
%A controlled PUF could be used to limit the usage of a key to a duration, etc

%Ephemeral PUF key, one-time signature PUF key, TRNG-based key?
\section{Oblivious Root-of-Trust Usage}\label{sec:rot-usage}
In this section we wish to focus on the challenge of using the root-of-trust without it
being vulnerable to attacks. As discussed in the previous section, a PUF can be used to
generate a signing key in a secure way since the generation process of the key depends
on physical variations which probing would disturb and cause the key generation step to
produce a different key. Moreover, the key is only ever present in volatile memory when
the chip is powered up.
\textit{It's assumed that reading the volatile memory is infeasible or very hard.}
%Hence, the key cannot be read when the device is powered down, from a non-volatile memory location.
But we need to use the key to perform cryptographic
operations such as signing. Does using the key expose the key to attacks? For example,
TEEs such as Intel SGX and Sanctum, depend on a privileged enclave that has access to a
signing key that is derived from the root-of-trust. Neither Intel SGX nor Sanctum
protect the attestation (signing) key from side-channel attacks, such as power analysis
attacks\cite{sgx:sanctum:8187110}, and it therefore seems reasonable to assume that the
attestation key could be compromised by physical or side-channel attacks when it is
under use, such as signing attestation reports.

\subsection{Masking and redundancy} In this work, we wish to protect the root-of-trust
against side-channel and fault injection attacks. The state-of-the art technique to
protect against side-channel attacks is
masking\cite{hpc:masking:9190067,10.1007/3-540-48405-1_26,momin2022handcrafting,koblaheda,cryptoeprint:2024/891,tches-2021-30795},
meanwhile redundancy is used to protect against fault injection
attacks\cite{attacks:faultinjection:sorcerer,cryptoeprint:2023/1769}. Masking can be thought of as MPC in
silicon, such that a secret key is split into shares and the computation is done on
shares. The masking scheme ensures that side-channel attacks such as power analysis or
electromagnetic radiation analysis 
cannot collect enough sufficient information from the shares to reconstruct the secret key\cite{hpc:masking:9190067}.
Redundancy techniques run the same computation multiple times through the same circuitry
or through duplicate circuitry, and compare the results of the different runs to make sure
they are the same. Hence, if some fault was injected in one run it will be detected. Recent work shows
that the usage of prime-field masking\cite{masking:prime-field:cassiers} can help
protect against fault injection attacks as well\cite{masking:prime:moos:tches24}.
It should be noted that masking and redundancy provide higher security at a cost with respect to
performance.

\subsection{Security of Masking Schemes}
It is important to note that masking schems depend on a source of
randomness\cite{9889394}. A true random number generator (TRNG) can be used to securely
derive the randomness. TRNGs leverage the noise (thermal, jitter, chaos, metastability,
etc)\cite{acosta2017embedded,yang2018true} in a chip as a source of entropy to derive randomness, and
have security properties very simlar to that of PUFs, since the randomness is also
derived from unpredictable and uncontrollable physical properties of the chip's
circuitry. TRNGs can be attacked though, especially if not designed or implemented
properly. For a survey of attacks on TRNGs see\cite{chowdhury2021physical}. To mitigate
the risks associated with attacks on the TRNG, some additional techniques are
suggested in\cite{9889394}.

%\subsection{Masking: Side-Channel Attack Protection}
%
%\subsubsection{Boolean Masking}
%\subsubsection{Arithmetic Masking}
%\subsubsection{Prime-Field Masking}

%\subsection{Redundancy: Fault-Injection Attack Protection}
\section{Crypto-Physically Secure Computations}\label{sec:secure-computations}
Recall that the goal of a TEE is to perform computations meanwhile providing
integrity and confidentiality guarantees. Intel SGX and Sanctum use hardware isolation
to secure the computations\cite{sgx:sanctum:8187110}, but the threat model excludes
physical side-channel attacks, and fault injection attacks\cite{cryptoeprint:2023/1769}. The state-of-the-art
techniques to protect against SCA and FI attacks is
masking\cite{hpc:masking:9190067,10.1007/3-540-48405-1_26,momin2022handcrafting,koblaheda,cryptoeprint:2024/891,tches-2021-30795},
and redundancy\cite{attacks:faultinjection:sorcerer}, which were
both discussed in Section~\ref{sec:rot-usage}, for the purpose of protecting the
root-of-trust (signing key) when it is used for signatures. Masking and redundancy offer
security at a cost to performance, and it would be useful to better understand the
tradeoff and challenges\cite{10137330} especially if these techniques are to be used
for complex computations.

The work in\cite{ches:uhleanother} discusses the challenges and tradeoffs associated
with implementing masking in both software and hardware. In brief, hardware masking
can be designed to be provably secure, both in theory and practice, but cannot be
changed post-fabrication unless it is implemented on FPGAs. Software masking
can be changed after fabrication, but is not secure in practice, and has a high latency
overhead. To get the best of both worlds, a masked instruction set extension has been
proposed in\cite{gao2021instruction}, but was shown to have flaws
in\cite{ches:uhleanother}. More recent work presents another instruction set
extension\cite{masking:eliminate-ise} that aims at elminating leakage stemming from
architectural and microarchitectural overwriting. This approach is still experimental
and is a matter of debate as to whether it can meet its security
goals\cite{ches:uhleanother}. Verifying that the masking used does not leak information
is important and is studied in\cite{hadvzic2024quantile} in the context of large masked
computations.

Secure computations are likely to require memory to be
encrypted\cite{armknecht2010memory}.
%
%
%\paragraph{Secure Boot}
%According to\cite{neopuf:9440114} SRAM PUFs cannot be used to secure the boot of a
%system:
%\begin{quote}
%``Due to the instability of its design, an SRAM PUF usually relies on software
%post-processing algorithms to ensure the correctness of the PUF-generated random
%numbers. The integrity of this software algorithm, however, cannot be ensured before
%establishing a secure operating environment. In this sense, a reliable SRAM PUF data
%cannot be derived before secure boot is finished, and, therefore, an SRAM PUF is unable
%to secure the boot flow of a system''.
%\end{quote}

%\subsection{Masking and Instruction Set Extensions}
%%%%%%%%%%%%%%%%%%%%%%%%%%%%%%%%%%%%%%%%%%%%%%%%%%%%%%%%%%%%%%%%%%%%%%%%%%%%%%
%                         Open Source Hardware
%%%%%%%%%%%%%%%%%%%%%%%%%%%%%%%%%%%%%%%%%%%%%%%%%%%%%%%%%%%%%%%%%%%%%%%%%%%%%%
\section{Open Source Hardware}
The core reasoning behind moving from close source to open source hardware is that it would encourage collaboration, which would eventually lead to better hardware and at a faster pace. Consider the story behind the \href{https://ohwr.org/project/cernohl}{CERN Open Hardware License}\cite{oshw:cernohl}\cite{oshw:cern}:
\begin{quote}
``For us, the drive towards open hardware was largely motivated by well-intentioned envy of our colleagues who develop Linux device-drivers'', said Javier Serrano, an engineer at CERN's Beams Department and the founder of the OHR.``They are part of a very large community of designers who share their knowledge and time in order to come up with the best possible operating system. We felt that there was no intrinsic reason why hardware development should be any different''.
[\ldots]
``By sharing designs openly'' said Serrano, CERN expects to improve the quality of designs through peer review and to guarantee their users --- including commercial companies --- the freedom to study, modify and manufacture them, leading to better hardware and less duplication of efforts.
\end{quote}
For some reason, the hardware world does not embrace open source like the software world. Moreover, it is common practice to use hardware obfuscation\cite{shakya2017,Fyrbiak2018OnTD} as a core design principle to secure hardware. Simply said, for whatever reason, the current hardware industry appears to be dominated by the belief that it's best to hide the design and inner workings of a chip by adding unnecessary elements to the design, just to confuse a potential attacker, in the hope that the attacker will not be able to understand the design, and thus reverse engineer it. Not everyone agrees with this vision. The Secure Cryptographic Implementation Association (SIMPLE-Crypto)\cite{simplecrypto} has already established a very good foundation to research and develop open source hardware. They have implemented AES in hardware with strong side-channel security countermeasures, which is currently under public evaluation. 
%See the outline of their vision at https://www.simple-crypto.org/about/vision/ and a detailed description of how they operate at https://www.simple-crypto.org/about/organization/.
%\subsubsection{Kerckhoffs's Principle applied to Chip Design}\label{kerckhoffss-principle-applied-to-chip-design}
Auguste Kerckhoffs, back in 1883, in his paper entitled ``La Cryptographie Militaire'' (Military Cryptography), argued that security through obscurity wasn't a desirable defense technique.
\begin{quote}
\textbf{\emph{Il faut qu'il n'exige pas le secret, et qu'il puisse sans inconvénient tomber entre les mains de l'ennemi}}
\end{quote}
roughly translated to:
\begin{quote}
\textbf{\emph{It must not require secrecy, and it must be capable without inconvenience to fall into the enemy's hand}}
\end{quote}
(Perhaps, one may point out that Kerckhoffs was assuming that the private key would be held secretly and not be part of an open design.) The need to secure a private key in an open design begs for physics to enter the arena (e.g. PUFs). The Secure Cryptographic Implementation Association (SIMPLE-Crypto Association) aims to apply Kerckhoffs's Principle to hardware and lays out their vision\footnote{\href{https://www.simple-crypto.org/about/vision/}{www.simple-crypto.org/about/vision}}:
\begin{quote}
\ldots\textbf{our vision is that as research advances, the security by obscurity paradigm becomes less justified and its benefits are outweighted by its drawbacks.} That is, while a closed source approach can limit the adversary's understanding of the target implementations as long as their specifications remain opaque, it also limits the public understanding of the mechanims on which security relies, and therefore the possibility to optimize them. By contrast, an open approach to security can lead to a better evaluation of the worst-case security level that is targeted by cryptographic designs.
\end{quote}

\paragraph{Open source hardware development} depends on three key things\cite{oshw:ihp}:
\begin{itemize}
    \item Open Electronic Design Automation (EDA) sofware tools
    \item Open Process Design Kit (PDK) software
    \item Foundries to build the chip, that will allow the design to be open source
\end{itemize}

\subsection{Open Source EDA Tools}
Open source electronic design automation (EDA) sofware such as
OpenRoad\cite{kahng2021openroad,10.1145/3505170.3511479,9643553,oshw:ve-hep-openroad}
can be used to design chips and be sent for tapeout at foundries, such as Google
Skywater and IHP, that support open sourcing the design. An EDA workflow to optmize the
design of masking schemes is presented in\cite{koblaheda} and may be useful to consult.

\subsection{Foundries and Open Source PDKs}
The Leibniz Institute for High Performance Microelectronics (IHP) maintains an open
source PDK that targets a 130 nm process node (SG13G2) of their pilot line, which
 manufactures circuits using high-performance SiGe BiCMOS technologies\cite{10614043}.
Google currently maintains three open source PDKs\cite{google-pdks}, in partnership with two foundries:
SkyWater Technologies (90nm and 130nm) and
GlobalFoundries (180nm)\cite{ansell2020missing,edwards2020google,10608632,10203216,jaramillo2024automated}.
%,10658750}.

%\paragraph{Multi Project Wafer}\label{multi-project-wafer}
%
%\begin{itemize}
%\item
%  \href{https://developers.google.com/silicon}{Build Custom Silicon with
%  Google}
%\item
%  \href{https://efabless.com/open_shuttle_program}{efabless}
%\item
%  \href{https://www.skywatertechnology.com/technology-and-design-enablement/mpw-programs/}{SkyWater}
%\item
%  \href{https://opensource.googleblog.com/2022/10/announcing-globalfoundries-open-mpw-shuttle-program.html}{Google
%  funds open source silicon manufacturing shuttles for GlobalFoundries PDK}
%\end{itemize}

\subsection{Research and Community Initiatives}
\begin{itemize}
    \item SIMPLE-Crypto --- ``develops open-source implementations of
        cryptographic algorithms, specialized for embedded systems (hardware and software),
        with strong physical security guarantees (e.g., against side-channel and fault attacks),
        featuring state-of-the art security and performance, maintained in the long-term''\cite{simplecrypto}
    \item The HEP Alliance\cite{hep-alliance, hep-press-release} is dedicated to ``Hardening the value
        chain through open source, trustworthy EDA tools and processors'', and is 
        working on a project (VE-HEP) to demonstrate the feasibility of using open
        source tools to develop secure chips.
\begin{quote}
    ``The use case to implement all these ambitious goals is the design and fabrication of an
    open-source hardware security module (HSM) that will be integrated into an automotive
    application''.
\end{quote}
    \item Free Silicon Foundation (F-Si)\cite{fsic2024}
    \item The FOSSi Foundation --- the custodian of the Free and Open Source Silicon movement\footnote{\url{https://fossi-foundation.org}}
    \item Workshop on Open-Source EDA Technology --- The WOSET workshop aims to galvanize the open-source EDA movement\footnote{\url{https://woset-workshop.github.io/}}
    \item The Silicon Salon --- Semiconductor Solutions for Cryptography\cite{silicon-salon} 
\end{itemize}

\subsection{Economic Challenges for Open Source Hardware}
What is the economic model for open source hardware?

\subsection{Technical Challenges for Open Source Hardware}
Open source hardware is currently far behind the cutting-edge closed source hardware.
What can be done to work towards closing that gap? See for
instance\cite{marshall2019hardware}, which discusses the gap between commercial and
open source hardware tools.

%\paragraph{Resources}\label{resources}

%\href{https://tinytapeout.com}{Tiny Tapeout} has a lot of educational material at that may be worth reading for those who don't have a background in hardware.

%\subsection{Continuous Public Evaluations}
%Public framework for continuous cryptanalysis to attempt breaking all aspects of the chip. Given its ambitious goals, the chip should be subject to extensive tests by security researchers, who would make their findings public and transparent with respect to their methodology, etc. These could be logged on a public ledger.
%%%%%%%%%%%%%%%%%%%%%%%%%%%%%%%%%%%%%%%%%%%%%%%%%%%%%%%%%%%%%%%%%%%%%%%%%%%%%%%%%%%%%%%%%
%               Proof of Fabrication (Verifiable Chip Implementation)
%%%%%%%%%%%%%%%%%%%%%%%%%%%%%%%%%%%%%%%%%%%%%%%%%%%%%%%%%%%%%%%%%%%%%%%%%%%%%%%%%%%%%%%%%
\section{Proof of Fabrication (Verifiable Chip Implementation)}\label{sec:proof-of-fab}
How do we know whether a given chip corresponds to a given design? Some possible
approaches, which can perhaps be combined together:

\begin{itemize}
    \item (\textbf{Pre-fab}) Logic Encryption encrypts the design\cite{Rajendran2017}
    \item (\textbf{Post-fab}) Microscope imaging of the chip to compare it against its design\cite{10179341}
    \item (\textbf{Trusted Supply Chain}) A public blockchain to track the stages of fabrication of a
        chip, which can be uniquely identified by its PUF or TRNG-based key\cite{10384393}
\end{itemize}

%%%%%%%%%%%%%%%%%%%%%%%%%%%%%%%%%%%%%%%%%%%%%%%%%%%%%%%%%%%%%%
\subsection{Pre-Fabrication: Logic Encryption}\label{pre-fabrication-logic-encryption}

Logic Encryption\cite{Tehranipoor2024,10483335,Rajendran2017} locks the
netlist of a chip design to protect
against a malicious foundry. The key to unlock the netlist is written to non-volatile
memory after fabrication, by the IP owner\cite{9300258}. The company HENSOLDT
Cyber\cite{hensoldt-cyber} has numerous research
works\cite{sisejkovicARCS2020,sisejkovicETS2019,sisejkovicVLSIDAT2019,sisejkovicSAMOS2018},
on the topic, in addition to actually making chips, and hence, is probably worth studying.
%Their papers are listed at https://hensoldt-cyber.com/scientific-papers/, but let's list a few here:

Logic encryption schemes can be attacked though. One type of attack targets the key used
to unlock the netlist\cite{9300258}, as the key is being transferred from memory to the
key-gates to unlock. The author in\cite{9300258} propose some countermeasures. Protecting the
locking key from physical attacks is a field of research. For instance, recent
 work\cite{10611730} propose to use nanomagnet logic to secure the locking key.

%In\cite{vanderleest:root-of-trust}, a PUF is used to
%generate the root of trust, which is used to identify a chip, and record its
%manufacturing phase and owner on a blockchain. Quoting the
%whitepaper\cite{vanderleest:root-of-trust}:
%\begin{quote}
%``Note that the blockchain technology can provide the trusted traceability
%along the supply chain. It tracks ``who owns what'' at each stage based on credentials protected by the
%embedded RoT. The PCB module manufacturers receive the certified RoT chips as components along with
%digitalized design parameters so that manual input can be avoided. Each time the medium products are
%handed off to the next stage, the “trust” is also transferred, and the credentials are verified and recorded
%in the blockchain''.
%\end{quote}

%See Root of Trust by Vincent Van der Leest et al.\footnote{https://www.synopsys.com/dw/doc.php/wp/gsa-end-to-end-traceability-of-ip-wp.pdf}.

%%%%%%%%%%%%%%%%%%%%%%%%%%%%%%%%%%%%%%%%%%%%%%%%%%%%%%%%%%%%%%
\subsection{Post-Fabrication: Microscope Imaging}\label{post-fabrication-microscope-imaging}
See\cite{10179341} in which SEM imaging was used to detect hardware trojan insertions in
chips. 

Some imaging techniques (invasive) destroy the chip in the process meanwhile others
(non-invasive) do not. Invasive analysis would need to be combined with a
Cut-and-Choose protocol as proposed by Miller in\cite{cut-and-choose}\footnote{See comment at \href{https://github.com/sbellem/qtee/issues/2\#issuecomment-1464600086}{https://github.com/sbellem/qtee/issues/2\#issuecomment-1464600086}}.

It's important to point out that there seems to be newer techniques that are
non-invasive, based on X-ray ptychography, X-ray nanotomography\cite{img:karpov2024high}
or Photonic Emission Analysis/Microscopy (PEM).

Quoting\cite{10179341}:

\begin{quote}
New non-invasive scanning methods based on X-Rays\cite{holler2019three} seem more promising for the future than the lengthy process of delayering and imaging the chip. These non-invasive techniques are potentially able to scan all metal layers and provide a 3D-image of the entire routing without destroying the device, but the research on this subject is still at an early stage.
\end{quote}

Also see\cite{cryptoeprint:2023/075} and\cite{img:karpov2024high,holler2019three},
and\cite{huang2023infraredinsituirisinspection}.

%X-ray nanotomography
%%%%%%%%%%%%%%%%%%%%%%%%%%%%%%%%%%%%%%%%%%%%%%%%%%%%%%%%%%%%%%
\subsection{Blockchain and PUF-based Trusted Supply Chain}\label{trusted-supply-chain}
Various works propose to leverage the use of a blockchain to trace the various steps
of the supply chain. These works make use
of a PUF or TRNG to uniquely identify a chip. See for
instance\cite{10384393,10570172,10.1145/3315571,9701029,8373269,10.1145/3315669,vanderleest:root-of-trust}.

\subsubsection{Blockchain Record of Verified Chips}\label{blockchain-puf-pubkeys}
As part of the trusted supply chain, the post-fabrication imaging
techniques discussed in Section~\ref{post-fabrication-microscope-imaging} could be
included, such that once a chip is verified its PUF public key could be logged into
a public blockchain, along with other relevant data.

\section{Unforgeable Attestation}
In the context of TEEs, attestation can be thought of as a proof provided by the
hardware that proves that a specific software binary is loaded into a specific
execution environment that will guarantee the integrity and confidentiality of the
computations, programmed in the software binary. Hence, they are two parts to the
attestation:

\begin{itemize}
    \item Proof of legit hardware
    \item Proof of loaded software binary
\end{itemize}

TEEs like Intel SGX\cite{cryptoeprint:2016/086} and
Sanctum\cite{sanctum:secure-boot:8429295}, implement the proof of legit hardware by using a
signing key that is tied to a public key associated with the manufacturer. The signing
key is secured by the hardware implementation and can only be accessed via a dedicated
enclave; (quoting enclave for Intel SGX, and signing enclave for
Sanctum)\cite{sgx:sanctum:8187110}. Neither
Intel SGX nor Sanctum protect against physical side-channel attacks, such
as differential power analysis (DPA) attacks, and it is therefore reasonable to assume
that an attacker with physical access could extract the signing key and forge
attestations (proofs)\cite{rfc9334}. Needless to say that this is problematic.

A design that would be capable to withstand physical attacks could make use of PUFs to
generate the signing key, as discussed in Section~\ref{sec:rot-gen}. In order to protect
the key against attacks when it is used to sign attestations, masking and redundancy
techniques could be used as discussed in Sections~\ref{sec:rot-usage}
and~\ref{sec:secure-computations}.

How can the hardware prove that it is legit, without relying on a trusted manufacturer?
Using PUFs helps to secure the key, but it does not help with proving that the hardware
is implemented as expected, unless the PUF key was derived from the entire circuitry of
the chip, such that any modification to any part of the chip would cause the PUF to
yield a different key. Alternatively, SpaceTEE\cite{DBLP:journals/corr/abs-1710-01430}
could be used to protect the hardware from malicious modification. The hardware could undergo a trusted verification phase as
discussed in~\ref{blockchain-puf-pubkeys} such that its PUF public key could be looked
up in a public blockchain record. If SpaceTEE is used, the pre-launch physical protection
measures would be applied as part of the trusted verification phase. The measurement
data (e.g., moment of inertia) would be stored on-chain along with the PUF public key.

%Strong PUFs, despite their shortcomings\cite{ieee:puf:strong:9971721}
Novel remote attestation schemes are presented in\cite{ra:puf::10168259,ra:puf:cryptoeprint:2021/602,ra:puf:242032,puf:rom:ra:van2023theoretical,9474324} and be worth reviewing in the context of this research. For a thorough
and formal treatment of remote attestation in the more general context of control flow attestation, we
recommend consulting\cite{sha2024controlflowattestationconceptssolutions}, which also
stresses the importance of considering physical attacks.

\newpage
\section{Roadmap: The Path Towards Crypto-Physically Secure TEEs}
In this section we wish to explore the possible small steps that can be taken towards
implementing a TEE that meets the challenges presented in Section~\ref{sec:challenges}.
Some of these steps may also be worked in parallel. For instance, it's reasonable to
envision making progress on the open source hardware front
(Challenge~\ref{challenge:oshw}), meanwhile making progress on verifying that a chip
was fabricated as expected (proof-of-fabrication Challenge~\ref{challenge:proof-of-fab}).
Perhaps each challenge can be worked on separately, thus allowing the formation of
specialized teams to research and develop towards improved TEEs. In the meantime, what
does that mean for Web3 projects who wish to implement applications relying on TEEs?
Must they wait for a new chip? Must they accept the flaws of current TEEs, and work
arround them? We think there may be a middle way, where it's possible to complement
current TEEs with emerging technologies like PUFs, thus resulting in a wide array of
novel hybrid models.  As a support to this argument, there are works that have already
proposed this such as\cite{zhang2024teamworkmakesteework}.
%Keystone enclave\cite{keystone:10.1145/3342195.3387532}, a fully open source software
%framework for TEEs, is aiming to be production ready\cite{keystone:status}, and could
%thus be a used, in combination with a chip that provides a secure root-of-trust.

%It would be useful to better understand Sanctum, what has been implemented, and whether
%it would be worth implementing parts of it or all of it in new chips.

%\input{future-work}
\section{Conclusion}
We have a lot of work to do!
\appendix
%%\section{}
%%\label{apx:}
%\include{cryptophysics}
%\include{attacks}
%\include{puf-appendix}

\ifnum\anon=0

\subsection*{Acknowledgements}
We thank Thorben Moos, Fran\c{c}ois-Xavier Standaert and Alex Obadia for valuable
feedback.
%We thank Thorben Moos, Fran\c{c}ois-Xavier Standaert for valuable feedback.

\fi

\newpage
\bibliographystyle{splncs04}
\bibliography{refs}

\end{document}